\documentclass[aps,twocolumn,prl]{revtex4-1}

\usepackage{newlfont}
\usepackage{color}           % color to the text \textcolor{red,green,blue}{text to be colored}
\usepackage{soul}            % strickthrough the text \st{text to be stricked} and \underline{text to be underlined}
\usepackage{amssymb}
\usepackage{amsfonts}
\usepackage{amsmath}
\usepackage{wasysym}
\usepackage{graphicx}
\usepackage{epsfig}
\usepackage{amsthm}
\usepackage{bm}
\usepackage[colorlinks=true, citecolor=blue, urlcolor=blue ]{hyperref}

\begin{document}

%\title{Genuine Multisite Entanglement suppresses Multiport Classical Information Transfer}
\title{Genuine Multiparty Quantum Entanglement Suppresses Multiport Classical Information Transmission}
%\title{Genuine Multiparty Entanglement suppresses Multiport Classical Information Transmission}
%\title{Multiport Quantum Dense Coding Capacity and Genuine Multisite Entanglement}

\author{R. Prabhu, Aditi Sen(De), and Ujjwal Sen}

\affiliation{Harish-Chandra Research Institute, Chhatnag Road, Jhunsi, Allahabad 211 019, India}

\begin{abstract}
We establish a universal complementarity relation between the capacity of classical information transmission by employing a multiparty quantum state as  a multiport quantum channel, and the corresponding genuine multipartite entanglement. The classical information transfer is from a sender to several receivers by using the quantum dense coding protocol with the multiparty quantum state shared between the sender and several receivers. The relation is derived when the multiparty entanglement is the relative entropy of entanglement as well as when it is the generalized geometic measure. The relation holds for arbitrary pure or mixed quantum states of an arbitrary number of parties in arbitrary dimensions. 
%We subsequently derive the relation when the entanglement measure is the generalized geometric measure.
\end{abstract}
%
%In multipartite scenario, for arbitrary states -- both pure and mixed states -- we prove that there exists a complementary relation between capacity of a classical information transmission through a quantum channel, in the form of dense coding, and genuine multipartite entanglement measures. Such duality is applicable for arbitrary number of parties in arbitrary dimensions. We choose relative entropy of entanglement and generalized geometric measure as the genuine multipartite entanglement measures. Moreover, we find that for the arbitrary pure three qubit states, there always exists a one-parameter family of states by which the area between the generalized geometric measure and capacity of classical information transmission is bounded. 
%

\maketitle

%\section{Introduction}

The discoveries of the quantum communication strategies,
% and further work in that direction over last two decades, 
beginning with the protocols of  quantum dense coding \cite{BW}, quantum teleportation \cite{teleportation} and quantum key distribution \cite{BB84}, have revolutionized the way we think about modern communication schemes. The importance of such protocols lies in the fact that they can efficiently transmit classical or quantum information in a way that is better than what is possible by using classical protocols. Such quantum communication schemes have already been realized experimentally, and in particular, experimental quantum dense coding have been reported in several systems including photonic states, ion traps, and nuclear magnetic resonance
% and  can be realized in the laboratory over around 100 kms 
\cite{exp}. 
The protocols were initially introduced for communication between two separated parties. This is true for the theoretical discussions of these protocols as well as for the experimental demonstrations of the same. 
%party systems. 
It has already been established that in the case of such bipartite communication schemes between a sender and a receiver, shared quantum correlations 
%in the quantum state shared between the sender and the receiver 
play a key role.
% in the success of these quantum protocols. 

Commercialization of these protocols demand the implementations of these protocols in a multipartite scenario \cite{comm-review}. 
Classical information transmission, in the form of quantum dense coding, has already been introduced 
in  multiparticle systems \cite{dcamader}. Further work in this direction include Refs. \cite{dense-further}.
%, DCmonogamy}. 
%Recently, an exclusion principle for quantum dense coding protocol has also been established.
%, where it was demonstrated that 
%only one of the pair in a multipartite quantum states can have quantum advantage in dense coding. 
%It is natural to ask whether such advantage is related to the multipartite entanglement measure. 

In this paper,  we consider the multiport quantum channel for transmitting classical information where a sender wishes to send classical information individually to several receivers by employing the quantum
%\textcolor{red}{pseudo multiport} 
dense coding protocol with a multiparty shared quantum state between all the partners. We find that the quantum advantage in this quantum dense coding scheme is suppressed by the 
% of one of the pair of the multiparticle state is always suppressed by 
 genuine multipartite entanglement of the shared multiparty quantum state. 
 %\textcolor{blue}{
Let us stress here that the quantity, coherent information, which quantifies quantum advantage in the dense coding protocol can also be used to study quantum error correction \cite{Schumacher1996}, ``partial'' quantum information \cite{HOW2005}, and distillable entanglement \cite{Devatak2005}.
%  Moreover, it appears, up to an additive constant, after we maximize the Holevo quantity \cite{}, over a relevant set of operations performed. The Holevo quantity is in turn the asymptotically achievable rate of classical communication over quantum channels.
%The capacity of multiport quantum 
%\textcolor{red}{dense coding} \textcolor{blue}{information protocols} has a complementarity with the amount of shared genuine multisite entanglement. 
The complementarity is first demonstrated by using the relative entropy of entanglement \cite{VPRK}, and holds for arbitrary (pure or mixed) quantum states of an arbitrary number of parties and in arbitrary dimensions. To show that the  complementarity is potentially generic, we go on to demonstrate the complementarity for an independent measure of genuine multisite entanglement, the generalized geometric measure \cite{GGM}, in which case the relation again holds for pure as well as mixed multisite quantum states of an arbitrary number of parties in arbitrary dimensions. 
%\textcolor{red}{We also show that the latter relation again  holds for mixed states of an arbitrary number of parties whose two party density matrices are of rank three or lower.}
%\textcolor{blue}{
Due to the monogamy of quantum correlations \cite{CKW}, one intuitively expects that a high multipartite entanglement of a  quantum state will suppress the  reduced bipartite entanglements, and hence will reduce the capacity of multiport dense coding. Since a quantitative statement of the constraint on the multiparty entanglement due to the monogamy of bipartite entanglement is as yet missing, it is not straightforward to relate the monogamy of bipartite quantum correlations with a quantum advantage of multiport channel capacities. The complementarity relation derived in this paper can shed light towards a quantitative understanding in this direction.
%of three qubits. 
%Quantitatively, we show a complementary relation of advantage of classical information transmission of arbitrary multipartite quantum state (pure and mixed) in arbitrary dimension with the genuine multipartite entanglement measures. 

%The rest of the paper is presented  as follows. Sections \ref{sec-dense} and \ref{sec-multi-measure} present the definitions required in the paper. In Sec. \ref{sec-dense}, we formally describe the dense coding capacity and the corresponding
% quantum advantage 
%over its classical counterpart. In Sec. \ref{sec-multi-measure}, we provide brief definitions of the genuine multisite entanglements used. The results are presented in Sec. \ref{sec-results}. We provide a conclusion in Sec. \ref{sec-conclu}. 

%\section{Quantum dense coding capacity and the quantum advantage}
%\label{sec-dense}

\textit{Quantum dense coding capacity and the quantum advantage.} Quantum dense coding (DC) is a quantum communication protocol by which one can transmit classical information encoded in a quantum system from a  sender to a receiver \cite{BW}. The 
available resources for the transmission are a shared quantum state and a noiseless quantum channel to transmit the sender's part of the shared quantum state to the receiver's end.
If the sender, called Alice, and the receiver, called 
Bob,  share a bipartite quantum state
 \(\varrho_{AB}\), then the amount of classical information (in bits) that the sender can send to the receiver is given by \cite{Holevo, dcamader,dcgeneral}
%%\begin{equation}
%\label{eq:capdefi}
% {\cal C}_{AB} \equiv  
\({\cal C}(\varrho_{AB}) = \max\{\log_2 d_A, \log_2 d_A + S(\varrho_B) - S(\varrho_{AB})\}\),
%\end{equation}
where \(d_A\) is the dimension of Alice's Hilbert space and \(\varrho_B\) is the local density matrix of Bob's subsystem. 
\(S(\cdot)\) denotes the von Neumann entropy of its argument and is defined as \(S(\sigma) = -\mbox{tr} (\sigma \log_2 \sigma)\), for an arbitrary quantum state \(\sigma\). 
The capacity ${\cal C}(\varrho_{AB})$ 
reaches its maximum when Alice and Bob share a maximally entangled state, which is a pure state with completely mixed local density matrices. 
Without using the shared quantum state but using the  noiseless quantum channel, Alice will be able to send \(\log_2 d_A\) bits. This process of sending classical information without using the shared quantum state is referred to as the ``classical protocol''.
 Using the shared 
quantum state is therefore advantageous if  \(S(\varrho_B) - S(\varrho_{AB}) > 0\).
A bipartite quantum state is said to be dense-codeable in that case. Correspondingly, the quantity  
%\begin{equation}
${\cal C}_{adv}(\varrho_{AB}) = \max\{S(\varrho_B) - S(\varrho_{AB}), 0\}$
%\end{equation}
 is identified as the ``quantum advantage'' in a dense coding protocol from Alice to Bob.  Note that 
%\begin{equation}
 \({\cal C}(\varrho_{AB}) = \log_2 d_A + {\cal C}_{adv}(\varrho_{AB})\).
%\end{equation}

Let us now move on to a multiport situation and suppose now that there are \(N+1\) parties who share a quantum state \(\varrho_{A B_1 B_2 \ldots B_N}\). Moreover, we assume that
among the \(N+1\) parties, \(A\) is the sender and the others, i.e \(B_1, B_2, \ldots, B_N\)  are the receivers. Let us consider a situation where \(A\) wants to send, individually, classical information
to the \(B_i\)'s \((i=1,2, \ldots ,N)\). 
In this multiport case, the quantum advantage is naturally defined as 
%
%, then it was shown recently that \(A\) can get quantum advantage in dense coding protocol at most one of the channel \cite{DCmonogamy} between the sender \(A\) and the 
%receiver, say \(B_i\). We define the
%advantage of dense coding protocol as
\begin{equation}
 {\cal C}_{adv}^{max}(\varrho_{AB_1B_2\ldots B_N}) = \max \{{\cal C}_{adv}(\varrho_{AB_i})|i=1,2,\ldots, N\},
% \label{eq:DCadv}
\end{equation}
where \(\varrho_{AB_i}\) is the local density matrix of \(A\) and \(B_i\) (\(i=1,2,\ldots,N\)).
This 
%``\textcolor{red}{pseudo} 
``multiport dense coding quantum advantage'' quantifies the amount of classical information that can be sent from the sender \(A\) to the \(N\) receivers by using the 
quantum dense coding protocol with the shared quantum state \(\varrho_{AB_1 B_2 \ldots B_N}\), over and above the amount of classical information that can be sent by using a classical protocol. 
%\textcolor{red}{
It is to be noted that the multiport dense coding protocol under consideration requires a multiparty state, and the successful implementation of the protocol will take place between the sender and a particular receiver, with the 
choice of the receiver (from among the many available) being  dependent upon the multiparty shared state that is considered as the channel for the said protocol.

%\section{Genuine Multipartite Entanglement Measures}
%\label{sec-multi-measure}

%In this section, 

\textit{Genuine multipartite entanglement measures.} We now introduce two multipartite entanglement measures which can quantify genuine multiparticle entanglement of an arbitrary quantum state \(\varrho_{A_1 A_2 \ldots A_n}\)
shared between \(n\) parties.

%\subsection{Genuine Multisite Relative Entropy of Entanglement}
\textit{Relative entropy of entanglement} --
The relative entropy of entanglement was proposed as a measure of entanglement for an arbitrary  multipartite state, $\varrho_{A_1 A_2 \ldots A_n}$  \cite{VPRK} as 
%\textcolor{blue}{For bipartite state, \(\varrho_{AB}\), it} is given by 
%%\begin{equation}
%\(E_R(\varrho_{AB}) = \min_{\sigma \in \mbox{\scriptsize{sep}}} S(\varrho||\sigma)\). 
%%\label{eq:REdefi}
%%\end{equation}
%Here, ``sep'' is the set of all separable states in \(A:B\),
% and the relative entropy, \(S(\varrho||\sigma)\), between \(\varrho\) and \(\sigma\), 
%is defined as \(S(\varrho||\sigma) = \mbox{tr} (\varrho \log_2 \varrho - \varrho \log_2 \sigma)  \).  
%It was shown that the relative entropy of entanglement satisfies all the properties which are required of a ``good'' entanglement %measure \cite{VPRK}.
%\(E_R\) reaches its maximum for a maximally entangled state. 
% -- nonvanishing for entangled state and non increasing under 
%local operations and classical communication. Moreover, the measure was shown to lie between two other entanglement measures, distillable entanglement \cite{HHHH-RMP} and entanglement of formation \cite{VPPRA}. 
%
%In a multipartite scenario, to quantify genuine multipartite entanglement of arbitrary multipartite quantum (pure or mixed) states, the %relative entropy of entanglement \textcolor{blue}{has been} defined as \cite{VPRK}
%\begin{equation}
\[E_R(\varrho_{A_1 A_2 \ldots A_n}) = \mathop{\min_{\sigma \in \mbox{\scriptsize{n-gen}}}} S(\varrho_{A_1 A_2 \ldots A_n}||\sigma),\] 
%\label{eq:relentropy}
%\end{equation}
where ``n-gen'' is the set of all n-party multipartite quantum (pure or mixed) states  which are not genuinely multipartite entangled, and the relative entropy, \(S(\varrho||\sigma)\), between \(\varrho\) and \(\sigma\), 
is defined as \(S(\varrho||\sigma) = \mbox{tr} (\varrho \log_2 \varrho - \varrho \log_2 \sigma)  \). 
An n-party quantum state is said to be genuinely multiparty entangled if it cannot be written as a probabilistic mixture of multiparty quantum states which are separable across at least one bipartition of the \(n\) parties. As an example, for three-party quantum systems between \(A_1\), \(A_2\), and \(A_3\), a probabilistic mixture of two quantum states which are respectively separable across \(A_1:A_2A_3\) and \(A_2:A_1A_3\) is not genuinely multisite entangled. \(E_R\) is  the ``relative entropy distance'' of the corresponding multisite state from the convex set of all multipartite states which are not genuinely multiparty entangled. 

%\subsection{Generalized Geometric Measure}
\textit{Generalized geometric measure (GGM)} --
The generalized geometric measure \cite{GGM} was first proposed to be a measure of genuine multipartite entanglement of a pure multiparty quantum state, by using a distance function of the given pure state from all pure multisite
 states which
are not genuinely multipartite entangled, and is defined as 
%\begin{equation}
\[{\cal E}(|\psi\rangle_{A_1 A_2 \ldots A_n}) = \min (1 - |\langle \phi|\psi\rangle|^2),\]
%\label{eq:GGMdefi}
%\end{equation}
where the minimization is over all \(|\phi\rangle_{A_1 A_2 \ldots A_n}\) that  are not genuinely multipartite entangled. 

%\st{This definition can be extended to arbitrary multiparty mixed quantum states
%In this paper, we extend the generalized geometric measure for arbitrary mixed states 
%by using the convex roof approach. 
%Therefore, the generalized geometric measure for an arbitrary mixed quantum state can be defined as}
%\textcolor{red}{
Using the convex roof approach, the generalized geometric measure for an arbitrary mixed quantum state can be defined as
\(
%\begin{equation}
{\cal E}(\varrho_{A_1 A_2 \ldots A_n}) = \min \sum_i p_i {\cal E}(|\psi_i\rangle_{A_1 A_2 \ldots A_n})\),
%\end{equation}
where the minimization is performed over all pure state decompositions of \(\varrho_{A_1 A_2 \ldots A_n} = \sum_i p_i (|\psi_i\rangle  \langle \psi_i|)_{A_1 A_2 \ldots A_n} \).

%\section{Complementarity}
%% between advantage of DC and GGM for entanglement measures}
%\label{sec-results}

%In this section, we 
\textit{Complementarity.} We now establish complementarity relations between the amount of classical information that can be sent through the multisite quantum state \(\varrho_{AB_1B_2 \ldots B_N}\), as quantified by the  
%\textcolor{red}{pseudo} 
multiport dense coding quantum advantage (\({\cal C}_{adv}^{max}\)),
%, as defined in Sec. \ref{sec-dense}, 
with genuine multipartite entanglement measures -- relative entropy of entanglement and generalized geometric measure.

%We will show such relation holds for arbitrary mixed states in arbitrary dimensions and for arbitrary number of parties.

%\subsection{Multiport dense coding advantage vs relative entropy of entanglement}

\textit{Multiport dense coding advantage vs relative entropy of entanglement} --
Let \(A\), \(B_1\), \(B_2\),..., \(B_N\) be \(N+1\) observers
%\textcolor{red}{
sharing an arbitrary $(N+1)$-party (pure or mixed) quantum state, $\varrho_{AB_1B_2\ldots B_N}$,  of arbitrary dimensions.
%\st{ who share the state  $\varrho_{AB_1B_2\ldots B_N}$, which is an arbitrary $(N+1)$-party (pure or mixed) quantum %state of arbitrary dimensions. }
In the 
%\textcolor{red}{pseudo} 
multiport  dense coding protocol that we consider, we assume that \(A\) is  the sender and \(B_i\)'s (\(i=1,2,\ldots,N\)) are the receivers. We now prove the following complementarity between the quantum advantage in this multiport scenario with the relative entropy of entanglement.

%, where $A$ is the sender and the $B_i$'s are the receivers. 
%By applying the definition of quantum advantage in dense coding, given in Eq. (\ref{eq:DCadv}) and relative entropy of entanglement, given in Eq. (\ref{eq:REdefi}), we prove the following theorem.
%Let \({\cal C}_{AB}\) and \({\cal C}_{AC}\) denote respectively the dense coding capacities of \(\rho_{AB}\) and \(\rho_{AC}\). 
%As was shown recently in Ref. \cite{amaderDcmonogamy}, the advantage of dense coding  by using quantum states over classical protocols %is possible only in one of the local states 
%of $\rho_{AB_1B_2\ldots B_N}$. Let us denote  the advantage of the capacity over classical protocol by \({\cal C}_{adv}\). The genuine %multipartite entanglement of $\rho_{AB_1B_2\ldots B_N}$
%is denoted by \({\cal E}\). Using these notations we now state the complementary relation between dense coding capacity and the genuine %multipartite entanglement. 

\noindent {\bf Theorem 1:} \emph{For the arbitrary multipartite pure or mixed quantum state \(\varrho_{AB_1B_2\ldots B_N}\) in arbitrary dimensions, 
the relative entropy of entanglement and the  
%\textcolor{red}{pseudo} 
multiport dense coding quantum advantage satisfy 
%, $E^G_R(\rho_{AB_1B_2\ldots B_N})$, and advantage in dense coding capacity 
%is bounded above by $\log_2 d$, where $d$ is the dimension of the receiver in the channel which has quantum advantage in dense coding, i.e.,
\begin{equation}
{\cal C}_{adv}^{max} + E_R \leq \log_2 d, 
\end{equation}
where \(d\) is the maximal dimension of the Hilbert spaces of the \(B_i\)'s.} 

\noindent \texttt{Proof.}   
%As it is defined in Eq. (\ref{eq:relentropy})
%We have
%The genuine relative entropy of entanglement is
One can obtain an upper bound on the multiparty relative entropy as follows
\begin{eqnarray}
&&E_R(\varrho_{AB_1B_2\ldots B_N}) = \mathop{\mbox{min}}_{\sigma \in \mbox{\scriptsize{(N+1)-gen}}} S(\varrho||\sigma)\nonumber \\
%\end{equation*}
%where $M$ is the set of all states ($\sigma$) which are not genuinely multipartite entangled. It is well known that the 
&\leq& \mathop{\min}_{\sigma\in \mbox{\scriptsize{sep}}_1} S(\varrho||\sigma') 
\equiv E^{AB_1:\mbox{\scriptsize{rest}}}_R(\varrho_{AB_1:B_2\ldots B_N})\nonumber\\
&\leq& E^{AB_1:\mbox{\scriptsize{rest}}}_f(\varrho_{AB_1:B_2\ldots B_N}) 
\leq S(\varrho_{AB_1}).
\label{entropyineq}
\end{eqnarray}
Here the set ``\(\mbox{sep}_1\)'' is the set of quantum states of \(A, B_1, B_2, \ldots, B_N\) which are separable across $AB_1:B_2\ldots B_N$.
\(E^{AB_1:\mbox{\scriptsize{rest}}}_R(\varrho_{AB_1:B_2\ldots B_N})\) and 
  $E^{AB_1:\mbox{\scriptsize{rest}}}_f(\varrho_{AB_1:B_2\ldots B_N})$ respectively denote 
the relative entropy of entanglement and the entanglement of formation \cite{Wooters} of the state \(\varrho_{AB_1B_2\ldots B_N}\) in the same bipartition.
%and $S(\varrho_{AB_1})$ is the von-Neumann entropy of the reduced state $\rho_{AB_1}$. 
The second inequality is obtained due to the fact that the relative entropy of entanglement is bounded above by the entanglement of formation \cite{VPPRA}, while the third follows from the fact that the von Neumann entropy of the local density matrix is an upper bound of  the entanglement of formation for bipartite states \cite{IBM-motka}.
%
% is obtained since the upper bound of  the entanglement of formation is the von Neumann entropy of local density matrix of \(\varrho_{AB_1}\) of 
%\(\varrho_{AB_1B_2\ldots B_N}\). 
% is the composite state between sender $A$ and the receiver $B_1$ who share the quantum channel which has dense coding advantage.

For the multipartite state $\varrho_{AB_1B_2\ldots B_N}$, the dense coding advantage can be written as
%\begin{eqnarray}
%{\cal C}_{adv}^{max}(\varrho_{AB_1B_2\ldots B_N}) &=& \max \{S_{B_1}-S_{AB_1}, S_{B_2}-S_{AB_2}, \nonumber \\
%& &\hspace{1.5cm} \ldots, S_{B_N}-S_{AB_N}, 0\}.\,\,\,\,\,\,\,\,\,
%\label{DCadvantage}
%\end{eqnarray}
${\cal C}_{adv}^{max}(\varrho_{AB_1B_2\ldots B_N}) = \max \{S_{B_1}-S_{AB_1}, S_{B_2}-S_{AB_2},
\ldots, S_{B_N}-S_{AB_N}, 0\},$
where \(S_{B_{i}}= S(\varrho_{B_{i}}) \) and \(S_{AB_i} = S(\varrho_{AB_i})\)  are  the single-site and  two-site von Neumann entropies  of \(\varrho_{AB_1B_2\ldots B_N}\) respectively. 
Consider the instance when 
%Suppose, we choose
 $S_{B_1}-S_{AB_1}$ attains the maximum.
%, i.e., $A$ and $B_1$ share the quantum channel through which they can successfully perform the dense coding, 
Then 
%\begin{equation}
\({\cal C}_{adv}^{max}(\varrho_{AB_1B_2\ldots B_N})=S_{B_1}-S_{AB_1}\).
%\label{cadv}
%\end{equation}
Adding this relation with Eq. (\ref{entropyineq}),
% and the relation for \({\cal C}_{adv}^{max}(\varrho_{AB_1B_2\ldots B_N})\),
%(\ref{cadv}),  
we obtain
%\begin{eqnarray}
${\cal C}_{adv}^{max}(\varrho_{AB_1B_2\ldots B_N}) + E_R(\varrho_{AB_1B_2\ldots B_N}) \leq S_{B_1}\leq \log_2 d_{B_1}.$  
%\leq \log_2 d_{B_1}.\nonumber
%&\equiv& \log_2 d.\,\,
%\end{eqnarray}
If the maximum of ${\cal C}_{adv}^{max}$ is obtained for the pair, say \(AB_i\),  
%in  \({\cal C}_{adv}^{max}\), 
we have to choose the corresponding bipartition in the relative entropy of entanglement in Eq. (\ref{entropyineq}), 
and in that case, the complementarity relation will be bounded
above by the logarithm of the dimension of subsystem \(B_i\). Therefore, in general, we have 
%for $N+1$ parties, we have 
{\normalsize\(\max\{ \log_2 d_{B_1},   \log_2 d_{B_2}, \ldots,  \log_2 d_{B_N}\}=\log_2 d\)}, as 
the upper bound for the sum \({\cal C}_{adv}^{max} + E_R\). 
Hence the proof. \hfill $\blacksquare$

The complementarity relation which is established above clearly indicates that a high  genuine multipartite entanglement will  lower the advantage of the same state for transmitting classical information. The result is universal in the sense that it holds for an arbitrary pure or mixed quantum state of an arbitrary number of parties in arbitrary dimensions.
% transmission 
%through the quantum channel shared between the pair of  bipartite state of $N+1$-party state.
%For arbitrary pure three-party state, the capacity of dense coding and genuine multipartite entanglement follow the following %complementary relation:
%i.e.
%\begin{eqnarray}
% \frac{1}{\log_2 d}{\cal C}_{adv} + log_d {\cal E} \leq 1,.
%\end{eqnarray}
%where \(d\) is the dimension of the senders 
%\noindent \texttt{Proof.}  

%\subsection{Relation between dense coding advantage and GGM}
\textit{Relation between dense coding advantage and GGM} -
Towards showing that the obtained complementarity relation is generic, we consider another genuine multiparty entanglement measure, the generalized geometric measure.
%, defined in Sec. \ref{sec-multi-measure}. 
The relation is first proven for pure multiparty quantum states in arbitrary dimensions. For simplicity, we assume that the state lies in \(\left(\mathbb{C}^d\right)^{\otimes {N+1}}\), with arbitrary dimension \(d\). 
%We first give the duality relation of the advantage in quantum dense coding protocol involving $N+1$ parties with the generalized geometric measure for pure states in arbitrary dimensions. 
Subsequently we show that the relation also holds  for arbitrary mixed states, $\rho_{AB_1B_2\ldots B_N}$.
%, whose two-party reduced density matrices are of rank 3 or less. 

Consider an $(N+1)$-party pure state, $|\psi\rangle_{AB_1B_2\ldots B_N}$, which is employed by the sender \(A\) to perform dense coding with the receivers 
%Such a state can be used to perform the dense coding protocol in 
%which $A$ is the sender and 
$B_i$'s ($i=1,2,\ldots,N$).
% are the receivers. The generalized geometric measure, which quantifies the genuine multipartite entanglement, is given by ${\cal E}(|\psi_{AB_1B_2\ldots B_N}\rangle)$.

\noindent {\bf Theorem 2:} \emph{The sum of the advantage in dense coding  and the generalized geometric measure for the arbitrary pure state $|\psi\rangle_{AB_1B_2\ldots B_N}$ is bounded above by unity, i.e.,}
\begin{equation}
\frac{1}{\log_2 d}{\cal C}_{adv}^{max} + \frac{d}{d-1}{\cal E} \leq 1.
\end{equation}
%where $d$ is the maximum of the dimension of the receiver. 
%
%\noindent \textbf{Remark.} 
%Note that 
Here the factors $\frac{1}{\log_2 d}$ and $\frac{d}{d-1}$, respectively for dense coding advantage and GGM, are normalizations that make the individual terms to have maximal value as unity.

\noindent \texttt{Proof.} 
%The normalized form of dense coding advantage is obtained by dividing Eq. (\ref{DCadvantage}) by the maximum of that quantity i.e., $\log_2 d$ on both the sides. Here $d$ is the dimension of the receiver who has the quantum channel with dense coding advantage. 
%\begin{eqnarray}
%\frac{{\cal C}_{adv}(\rho_{AB_1B_2\ldots B_N})}{\log_2 d} &=& \frac{1}{\log_2 d}\mbox{Max}[S_{B_1}-S_{AB_1}, S_{B_2}-S_{AB_2}, \nonumber %\\
%& &\hspace{1.5cm} \cdots, S_{B_N}-S_{AB_N}, 0].
%\label{eq:}
%\end{eqnarray}
Let us  assume, without loss of generality, that the maximum in the  
%\textcolor{red}{pseudo} 
multiport quantum advantage in dense coding is attained for \(S_{B_1}-S_{AB_1}\), i.e.
%between $A$ and $B_1$ is maximum, i.e.,
%\begin{eqnarray}
\({\cal C}_{adv}^{max}(|\psi\rangle_{AB_1B_2\ldots B_N}) = S_{B_1}-S_{AB_1}\).
%\label{eq:normdcadvantage}
%\end{eqnarray}
We now note that the GGM for the state $|\psi\rangle_{AB_1B_2\ldots B_N}$ can be shown to be given by \cite{GGM}
%\begin{equation}
\({\cal E}(|\psi_{AB_1B_2\ldots B_N}\rangle)=1-\max\{\Lambda_j\}\),
%\frac{d}{d-1}{\cal E}\left(\sum_i{p_i |\psi_i\rangle\langle\psi_i|}\right),
%\label{eq:}
%\end{equation}
where $\Lambda_j$'s are the maximal eigenvalues of the local density matrices of all possible bipartite partitions of the state $|\psi\rangle_{AB_1B_2\ldots B_N}$. 
%Since the highest eigenvalue of a two-party reduced density matrix, $\rho_{AB_1}$ of $|\psi_{AB_1B_2\ldots B_N}\rangle$ will be one of the  eigenvalue corresponding to all possible partitioning of the state $|\psi_{AB_1B_2\ldots B_N}\rangle$, then we have
%\begin{eqnarray*}
%{\cal E}\left(\sum_i{p_i|\psi_i\rangle\langle\psi_i|}\right)&\leq& {\cal E}\left(\sum_i{p_i\mbox{tr}_{B_2B_3\ldots %B_N}|\psi_i\rangle\langle\psi_i|}\right)\\
%&\leq&{\cal E} \left(\sum_i{p_i|\psi_{AB_1}\rangle\langle\psi_{AB_1}|}\right)\\
%&\leq&1-\lambda_{AB_1}.
%\end{eqnarray*}
%Therefore we have
%Rewriting the above equation by using the highest Schmidt numbers  obtained in all the partitioning of the state $\rho_{AB_1B_2\ldots B_N}$, we get
%$${\cal E}(\rho_{AB_1B_2\ldots B_N})=\sum_i{p_i \frac{d}{d-1}\mbox{Min}(1-\lambda_i)}.$$
%Again suppose in all the possible partitioning of the state $\rho_{AB_1B_2\ldots B_N}$, if $\rho_{AB_1}$ gives the highest eigenvalue, then the above equation will become
Therefore, we have 
%\begin{equation}
\({\cal E}(|\psi\rangle_{AB_1B_2\ldots B_N})\leq 1-\lambda_{AB_1}\),
%\label{eq:normggm}
%\end{equation}
with $\lambda_{AB_1}$ being the maximum eigenvalue of the two-party reduced density matrix $\varrho_{AB_1}$ of \(|\psi\rangle_{AB_1B_2\ldots B_N}\). Adding the relations for \({\cal C}_{adv}^{max}(|\psi\rangle_{AB_1B_2\ldots B_N})\) and GGM, 
%From Eqs. (\ref{eq:normdcadvantage}) and   (\ref{eq:normggm}), 
we obtain
\begin{eqnarray*}
\frac{{\cal C}_{adv}^{max}(|\psi\rangle_{AB_1B_2\ldots B_N})}{\log_2 d} &+&\frac{d}{d-1} {\cal E}(|\psi\rangle_{AB_1B_2\ldots B_N})  \nonumber \\
& &\leq \frac{S_{B_1}}{\log_2 d}-\frac{S_{AB_1}}{\log_2 d} + \frac{d(1-\lambda_{AB_1})}{d-1}.
\end{eqnarray*}
%\textcolor{blue}{
To complete the proof, we have to show that the sum of the last two terms is non-positive. The sum of these two terms of the above equation is a convex function. Therefore, its maximum is attained at the extremal points, i.e. for $\lambda_{AB_1}=\frac{1}{d^2},\, 1$. When $\lambda_{AB_1}=1$, sum vanishes and at $\lambda_{AB_1}=\frac{1}{d^2}$, it is negative. The proof follows from the fact that $S_{B_1} \leq \log_2 d$.
\hfill $\blacksquare$

%\textcolor{red}{\st{The von Neumann entropy of} $\varrho_{AB_1}$ \st{is related to its highest eigenvalue as }
%$S_{AB_1} \geq \log_2(\frac{1}{\lambda_{AB_1}})$ \cite{Wehrl1978}. \st{Using this inequality, the relation reduces to }
%%rewrite the above expression as
%\begin{eqnarray*}
%& &\frac{{\cal C}_{adv}^{max}(|\psi\rangle_{AB_1B_2\ldots B_N})}{\log_2 d} + \frac{d}{d-1}{\cal E}(|\psi\rangle_{AB_1B_2 \ldots B_N})  %\nonumber \\
%& &\hspace{1cm} \leq \frac{S_{B_1}}{\log_2 d} + \frac{\log_2 \lambda_{AB_1}}{\log_2 d} + \frac{d(1-\lambda_{AB_1})}{d-1}.
%\end{eqnarray*}
%\"st{It can be shown that  the sum of the second and third terms in the right-hand-side of the above inequality is always negative, 
%for all} $d>2$ \st{and it vanishes for} $d=2$. \st{The proof follows from the fact that} $S_{B_1} \leq \log_2 d$. }
%, remaining term will be less equal to unity, hence the proof.

We now consider quantum states of  $N+1$ parties which are possibly  mixed, and are in arbitrary dimensions. For simplicity, we assume that the state is defined  on \(\left(\mathbb{C}^d\right)^{\otimes {N+1}}\), with arbitrary dimension \(d\). 
% to establish the complementary relation stated above.

\noindent{\bf Theorem 3:} \emph{For the arbitrary (possibly mixed) quantum state $\varrho_{AB_1B_2\ldots B_N}$, the sum of the 
%normalized 
quantum advantage in dense coding and the 
%normalized 
generalized geometric measure is bounded above by unity.}

\noindent \texttt{Proof.} 
The genuine multiparty entanglement measure, GGM, of $\varrho_{AB_1B_2\ldots B_N}$ is given by
%\begin{equation}
\({\cal E}(\varrho_{AB_1B_2\ldots B_N})=\sum_k p_k {\cal E}(|\psi_k\rangle\langle\psi_k|)\),
%\label{eq:}
%\end{equation}
where the ensemble  $\{p_k,|\psi_k\rangle\}$ forms the optimal pure state decomposition of the state $\varrho_{AB_1B_2\ldots B_N}$ for obtaining the minimum in the convex roof of the GGM. Suppose that $\lambda_{AB_1}^k$ is the maximum eigenvalue of the two-party reduced density matrix, of the parties \(A\) and \(B_1\), of the state $|\psi_k\rangle$.
% of  $\rho_{AB_1B_2\ldots B_N}$. 
Then we have
${\cal E}(\varrho_{AB_1B_2\ldots B_N}) \leq \sum_k p_k (1-\lambda_{AB_1}^k).$
%\textcolor{red}{For the cases when the states \(\varrho_{AB_i}\) are of at most rank three,} 
One can show that
%\begin{eqnarray}
${\scriptsize \frac{d}{d-1}{\cal E}(\rho_{AB_1B_2\ldots B_N})  \leq  \sum_k p_k \frac{S(\mbox{tr}_{B_2 \ldots B_N}|\psi_k\rangle \langle \psi_k|)}{\log_2{d}} \leq \frac{S(\varrho_{AB_1})}{\log_2{d}}.}$
%\end{eqnarray}
We have used concavity of the von-Neumann entropy to get both the inequalities. Now, along with the quantum advantage in dense coding, %defined in the Eq. (\ref{eq:DCadv}), 
and using the above inequality, we obtain
% GGM will become
\begin{eqnarray}
\frac{{\cal C}_{adv}^{max}(\varrho_{AB_1B_2\ldots B_N})}{\log_2 d} &+& \frac{d}{d-1}{\cal E}(\varrho_{AB_1B_2\ldots B_N})\nonumber\\
& & \leq \frac{S_{B_1}-S_{AB_1}}{\log_2{d}}+\frac{S_{AB_1}}{\log_2{d}} \leq 1.\,\,\,\,\,\,\,\,
\label{eq:}
\end{eqnarray}
Here we have assumed, without loss of generality, that the maximum 
%in the definition 
of ${\cal C}_{adv}^{max}$ is attained by the \(AB_1\) pair. 
%$\rho_{AB_1}$ of the $\rho_{AB_1B_2\ldots B_N}$. 
\hfill $\blacksquare$

\begin{figure}%
\includegraphics[width=0.4\columnwidth,height=0.12\textheight]{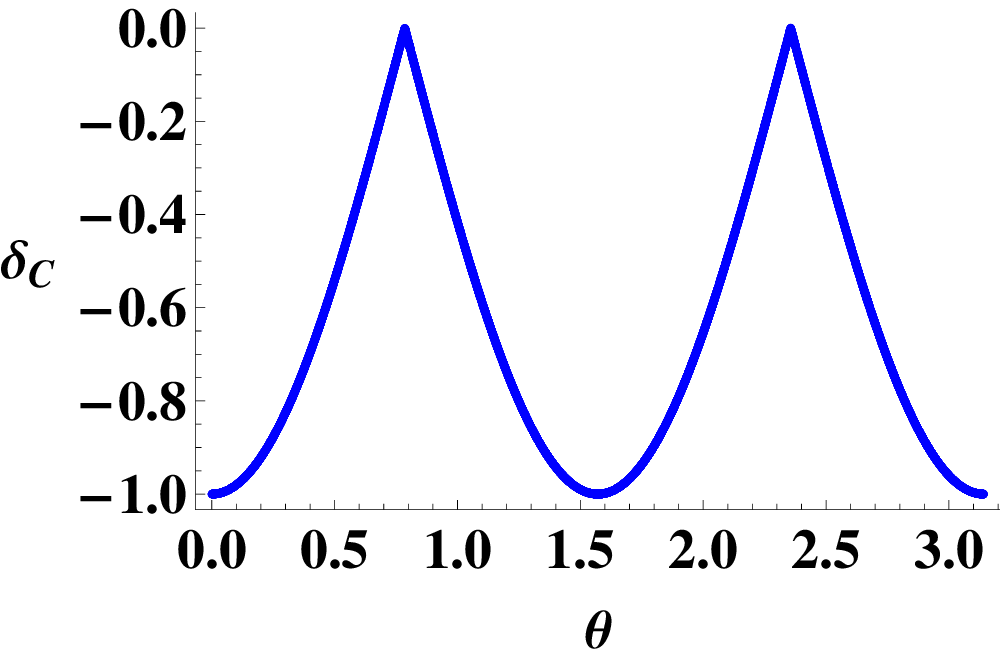}%
\includegraphics[width=0.58\columnwidth,height=0.15\textheight]{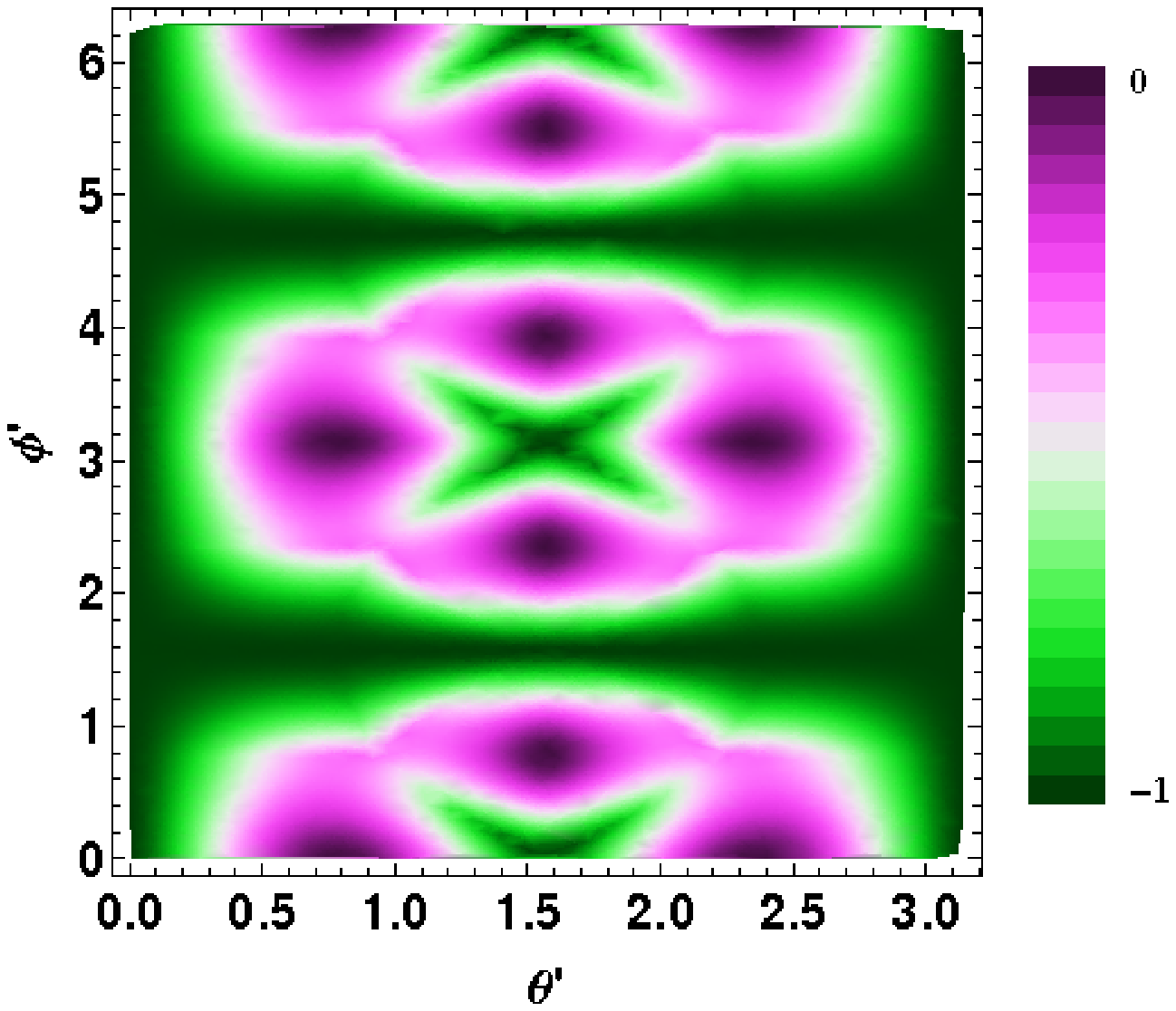}%
\caption{(Color online.) Saturation of the complementarity quantity.
The complementarity quantity, 
$\delta_C = {\cal C}_{adv}^{max} + {\cal E} -1 $  is plotted against the state parameter, $\theta$, of the generalized GHZ state (Left). When $\delta_C =0$, saturation of complementarity is achieved.
 The right figure presents the projection of 
%This is the projection plot of
the  $\delta_C$ surface, for the generalized W state, on the plane of the state parameters $\theta{'}$ and $\phi{'}$.
% of the generalized W state, given by
% \(|\psi_{W}\rangle = \sin\theta^{'}\cos\phi^{'}|011\rangle+\sin\theta^{'}\sin\phi^{'}|101\rangle+\cos\theta^{'}|110\rangle\).  
The dark violet regions show the area where $\delta_C$ vanishes, so that the complementarity relation is saturated. 
%It also signifies that in the state space, the complementary relation given in Theorem 2 is quite strict.  
The $\delta_C$ axis is dimensionless and all other axes are in radians.}
\label{fig:WGHZ}%
\end{figure}

We now illustrate that the obtained  complementary relation between the  
%\textcolor{red}{pseudo} 
multiport dense coding quantum advantage and genuine multipartite entanglement is tight.
For this investigation, 
%we consider  
% two important classes of tripartite pure quantum states of three qubits -- the generalized Greenberger-Horne-Zeilinger (GHZ) states \cite{GHZ}  and the generalized W states \cite{WDur}. 
let us first consider the generalized 
GHZ state \cite{GHZ},
%given by 
%\begin{equation}
\(|\psi_{GHZ}\rangle=\cos\theta|000\rangle+\sin\theta|111\rangle\),
%\end{equation}
where $\theta \in (0,\pi)$.
% and $\phi \in (0,2\pi)$. 
We plot the complementarity quantity
%\begin{equation}
\(\delta_C=\frac{1}{\log_2 d}{\cal C}_{adv}^{max} + \frac{d}{d-1}{\cal E}-1\)
%\end{equation}
with respect to
% the 
%state parameter
 $\theta$ 
(see Fig. \ref{fig:WGHZ}(left)). 
We see that the saturation of $\delta_C$ 
%the complementarity relation
 occurs at $\theta = \frac{\pi}{4}$ and $\frac{3\pi}{4}$.
% for all values of $\phi$.
Note here that the complementarity relation of Theorem 2 implies that \(\delta_C\) is always non-positive, and that the vanishing of \(\delta_C\) implies that the bound is tight.
%\begin{figure}%
%\includegraphics[width=0.5\columnwidth,height=0.15\textheight]{GHZgeneral2d}%
%\caption{(Color online.) 
%%The complimentary relation between quantum advantage in dense coding capacity  and generalized geometric measure have been pictorially produced, for better understanding of the strength of the relation. Here we consider 
%The complementarity quantity, 
%$\delta_C = {\cal C}_{adv}^{max} + {\cal E} -1 $  is plotted against the state parameter, $\theta$, of the generalized GHZ state. 
%%The saturation region (i.e. when $\delta_C= 0$) (dark violet) is observed at $\theta = \frac{\pi}{4}$ and $\frac{\pi}{2}$ for all values of $\phi$. 
%The horizontal axis is in radians, while the vertical axis is dimensionless.
%}%
%\label{fig:GHZ}%
%\end{figure}
Similarly, we consider the generalized W states \cite{WDur}, \(|\psi_{W}\rangle=\sin\theta{'}\cos\phi{'}|011\rangle+\sin\theta{'}\sin\phi{'}|101\rangle+\cos\theta{'}|110\rangle\),
%\end{equation}
with $\theta{'} \in (0,\pi)$ and $\phi{'} \in (0,2\pi)$, which are known to be inequivalent to the generalized GHZ states by stochastic local operations and classical communication \cite{WDur}.
%, given by
%\begin{equation}
%\(|\psi_{W}\rangle=\sin\theta{'}\cos\phi{'}|011\rangle+\sin\theta{'}\sin\phi{'}|101\rangle+\cos\theta{'}|110\rangle\),
%\end{equation}
%with $\theta{'} \in (0,\pi)$ and $\phi{'} \in (0,2\pi)$.
 In Fig. \ref{fig:WGHZ} (right), we  depict the projection of $\delta_C$ on the \((\theta{'}, \phi{'})\)-plane for the generalized W state. 
It clearly indicates that there are regions where 
%the complementarity relation obtained in Theorem 2 
\(\delta_C\) is saturated.

%\section{Conclusion}
%\label{sec-conclu}

Summarizing, we have established a complementary relationship between the quantum advantage of the 
%\textcolor{red}{pseudo} 
multiport classical capacity of a multiparty quantum state used as a quantum channel and the 
genuine multipartite entanglement of the same state. 
%We show that to obtain a quantum advantage over classical protocol of a quantum dense coding protocol of a state which is a local state of multipartite state, one should have substantially little genuine multipartite entanglement 
%in a multiparticle state. ??
The relation is demonstrated for two genuine multipartite entanglement measures -- the relative entropy of entanglement and the generalized geometric measure. The relation holds for pure or mixed quantum states of arbitrary dimensions 
and of an arbitrary number of parties. 
%The results strongly indicates that such complementarity is generic for arbitrary multipartite entanglement measures. ??
%coarse-graining!!

%\textcolor{red}{\st{Quantum dense coding between a sender and 
%%single 
%receiver is known to provide the maximal quantum advantage for the maximally entangled 
%%bipartite 
%quantum state. 
%Due to the monogamy of quantum correlations, one intuitively expects that a high multipartite entanglement of a  quantum state will %suppress the  reduced bipartite entanglements, and hence will reduce the capacity of multiport dense coding.
%%  for the multiparty state. 
%However, since a quantitative statement of the constraint on the multiparty entanglement due to the monogamy of bipartite entanglement %is as yet missing, it is not straightforward to relate the monogamy of bipartite 
%quantum correlations with a multiport quantum advantage of channel capacities. 
%%The complementarity relation derived is to be seen in this light. %???
%The complementarity relation derived in this paper can shed light towards a quantitative understanding in this direction.}}

%\acknowledgements 

R.P. acknowledges support from the 
%Department of Science and Technology, 
DST, Government of India, in the form of an INSPIRE faculty scheme at the Harish-Chandra Research
Institute, India.

\end{document}